\documentstyle[12pt,aps,floats]{revtex}

\newcommand{\bR}{{\bf R}}
\newcommand{\Ro}{R_{0}}
\newcommand{\bV}{{\bf V}}
\newcommand{\cV}{{\cal V}}

\newcommand{\re}{^{\rm e}}
\newcommand{\ri}{^{\rm i}}
\newcommand{\Om}{\Omega}
\newcommand{\Ga}{\Gamma}
\newcommand{\an}{\alpha}
\newcommand{\bn}{\beta}
\newcommand{\gn}{\gamma}

\newcommand{\ad}{_{\alpha}}

\newcommand{\bd}{_{\beta}}

\newcommand{\ano}{\alpha=1,2,3}

\newcommand{\pa}{\partial}
\newcommand{\Sum}{\sum\limits}
\newcommand{\Int}{\int\limits}
\newcommand{\Frac}[2]{\frac{\textstyle #1}{\textstyle #2}}

\newcommand{\reduction}[2]{\left. #1 \right|_{#2}}

\newcommand{\Bigcup}[1]{\mathop{\bigcup}\limits_{#1}}

\newcommand{\Lim}[1]{\mathop{{\rm lim}}\limits_{#1}}

\newcommand{\be}{\begin{equation}}
\newcommand{\ee}{\end{equation}}

\begin{document}
\title{FEW-BODY QUANTUM PROBLEM IN THE BOUNDARY-CONDITION MODEL%
\footnote{Talk given at the International
Workshop ``Meson--Barion interactions and Few--Body systems'',
Dubna (Russia), 28--30~April, 1994. 
LANL E-print {\tt nucl-th/9606022}.}}

\author{A.K.Motovilov}
\address{BLTP, Joint Institute for Nuclear Research, 
               Dubna, 141980, Russia\\
                 Email: motovilv@thsun1.jinr.dubna.su}
\maketitle
\bigskip

\bigskip

\begin{abstract}
Systems of three and four quantum particles in the boundary-condition model
are considered. The Faddeev--Yakubovsky approach is applied to
construct the Fredholm--type integral equations for these systems in
framework of the Potential theory.  The boundary--value problems are
formulated for the Faddeev--Yakubovsky components of wave functions.
\end{abstract}
\bigskip

\bigskip

\section{Introduction}

The report is concerned with a treatment of three- and four-body
quantum systems with pair interactions described by the boundary
conditions of various types.  Models of a such kind  including the
model of interactions with hard core attract attention  due to
simplicity of description of particle interaction at small distances
(see Refs. \cite{EfiSch},\cite{MMYa} for review). In these models,
the repulsive part of the pair interaction is described by the boundary
conditions for
wave function which are set on a surface $\partial\omega$
 in the (center of mass frame) two-body configuration space ${\bf R}^{n}$.
Here, $n$ is dimension of particles (usually $n=3$ and
$\partial\omega$ is a sphere in ${\bf R}^{3}$).

However, being really  simple in the two-body case,
the boundary condition model in the case of three and more particles
gives a birth to certain mathematical difficulties untypical for
few-body problems with smooth potentials. Thing is that one comes
here to necessity to construct a scattering theory in exterior of
noncompact surface formed
by aggregate of cylinders $\partial\omega_{\alpha}\times{\bf R}^{(N-2)n}$
being supports in the $N$-body configuration space ${\bf R}^{(N-1)n}$
for interactions in pair subsystems $\alpha$.
Standard equations of few-body scattering theory (see Ref.~\cite{MF})
were derived for non-singular interactions and are not adjusted to a
work with the arising boundary value problems.  There are few
attempts \cite{KimT}--\cite{Brayshaw} in physical literature to make
an immediate regularization of these equations in three-body case
using special limit procedures where at the beginning, one takes
regular potentials with finite repulsive cores and then the cores are
pulled to infinity.  In their results, the papers
\cite{EfiSch},\cite{VEfimov},\cite{KuKhar} are close to this
approach, too.  These attempts  were not quite successful because
resulting equations were not of Fredholm type. To make the problem
unique soluble it is necessary to take into account additional
considerations \cite{EfiSch},\cite{KuKhar}.

Another method is proposed in works \cite{MerMot}--\cite{Thesis} of
 S.P.Merkuriev and author. Following the traditional approach to
treatment of the boundary value problems for elliptic operators, we
use the Potential theory (see Ref.~\cite{Potential}) reducing the
problems to studies of integral equations on the surface where
boundary conditions are set on.
As mentioned above, this surface in the case of $N$-particle system with
$N\!\!\geq\! 3$ is unbounded. Therefore, in a difference to the boundary
value problems for compact surfaces, the equations of the Potential
theory are not of Fredholm type. To transform them into those of
Fredholm type  we apply in three-body case, the method by
L.D.Faddeev \cite{Faddeev63}. Namely we extract and inverse explicitly
 singular diagonal part of integral operator in the Potential
equations. As a result we obtain the Faddeev-type
equations for densities of simple and/or
double layer (kind of densities depends on a type of boundary
conditions) \cite{MerMot}--\cite{Thesis} which are did of Fredholm
type and make possible to ground the scattering problem
\cite{Thesis}. Note, in the recent works \cite{BG}, a new method to
prove the
completeness of wave operators in $N$-body problems with arbitrary $N$
has been developed.  This method is based on the concept of locally
conjugate operator and is extended on the hard-core $N$-body
Hamiltonians \cite{BGS}.

Alongside with the integral form we use also a differential form of the
Potential equations. Here, a very important thing are the generalized
potentials (quasipotentials) \cite{MMYa},\cite{Thesis} (one-dimensional
variants of quasipotentials for the boundary-condintion model were
constructed  for the first time in Refs. \cite{LMcM},\cite{H})
allowing to reformulate boundary conditions in terms of singular
distributions.  These potentials are presented by linear combinations
of the delta-function concentrated on the surface and its derivative
with respect to the surface normal.
The generalized potential method simplifies a scheme
of derivation of the Faddeev-Yakubovsky-type integral (in framework of
the Potential theory) as well as differential (in terms of
quasipotentials) equations for components of resolvent and then for
the wave operators. In the work \cite{MMYa} we formulate such
equations in the cases of three and four particles. However the
derivation may be spread also on the case of a system with arbitrary
number of particles. As in \cite{MF}, integral equations allow to
study asymptotical boundary conditions for the Yakubovsky
components of  wave functions at large values of space variables.
The formulations obtained of the boundary value problems for the
Faddeev--Yakubovsky differential equations may be applied to
study of concrete few-body systems in the boundary-condition model.
Some results of computations of three-body scattering and
bound-state energies  are presented in Refs. \cite{Motovilov},\cite{EChAYa}.

In the present work, we review the results from
Refs.~\cite{MerMot}--\cite{Thesis},\cite{MMYa} and \cite{EChAYa}
reformulating them for the case of arbitrary dimension
of particles $n\!\ge\! 2$.

\section{Notations}
In the report, we restrict ourselves only to the cases of $N\!\!=\! 3$
and $N\!\!=\! 4$ particles.
For the sake of simplicity we suppose all the particles
to be spinless.

For description of three-body system we shall use the standard reduced
relative coordinates \cite{MF} $x\ad,y\ad$, $\an=1,2,3$.
For example in the case of $\an=1$ these coordinates
are expressed through the radius-vectors $r_i\in\bR^n$ and
masses ${\rm m}_i$ of particles by the formulae
$$
x_{1}  =  \left[ \Frac{2{\rm m}_{2}{\rm m}_{3}}{{\rm m}_{2} + {\rm m}_{3}}
        \right]^{1/2}
        ({\mit r}_{2} - {\mit r}_{3}),\quad
y_{1}  =  \left[
              \Frac
               {2{\rm m}_{1}({\rm m}_{2} + {\rm m}_{3})}
             {{\rm m}_{1} + {\rm m}_{2} + {\rm m}_{3}}
        \right]^{1/2}
        \left(
              {\mit r}_{1} -
              \Frac
       {{\rm m}_{2}{\mit r}_{2} +
      {\rm m}_{3}{\mit r}_{3}}{{\rm m}_{2} + {\rm m}_{3}}
         \right).
$$
Relative coordinates are combined in $2n$-vectors
$X=(x\ad,y\ad)$. A choice of coordinate pair fixes cartesian coordinate
system in $\bR^{2n}$.

In the case of four-body system, alongside with the index $\an$
denoting again a pair subsystem, we use also the index $a$
for partition of the system in two subsystems. If the pair $\an$
belongs to one of these subsystems, we write $\an\subset a$.
As above, we introduce the relative coordinates $x\ad,y\ad$,
$x\ad\in\bR^n$, $y\ad\in\bR^{2n}$. Here, $y\ad$ is a set of relative
coordinates of the pair $\an$ considered as a whole and two particles
in the rest. A detailed description of relative coordinates for
four-body system can be found in Ref.~\cite{MF}.

In the boundary-condition model, the configuration space $\Omega$
of $N$-body system includes the points $X\in\bR^{(N-1)n}$
satisfying conditions $x\ad\in\omega\ad$
for all indices $\an$ where $\omega\ad$ ,
$\omega\ad \subset \bR^n$,  is the domain outside
with respect to a piece-wise smooth closed compact surface
$\partial\omega\ad \subset \bR^n$.
This $\partial\omega\ad $ is a surface where the boundary conditions
are set on in the problem of two particles belonging to pair $\an$.

By $\Gamma\ad$ we denote the $[(N-1)n-1]$--dimensional cylinders
in $\bR^{(N-1)n}$ engendered by surfaces $\partial\omega\ad$,
$\Gamma\ad=$ $\partial\omega\ad\times\bR^{(N-2)n}=$
$\{X\!\in\!\bR^{(N-1)n}:\, X=(x\ad,y\ad),\, x\ad\in\partial\omega\ad\}$.

Hamiltonian of $N$-body system is defined in $L_2(\Omega)$
by the expression
$$
Hf(X)=\left(-\Delta_X+\Sum_{\an}v\ad(x\ad)\right)f(X)
$$
on the functions $f\in W_2^2(\Omega)$ satisfying
the Dirichlet conditions (hard-core model)
\be
\label{bccor}
\reduction{f}{\partial\Omega}=0
\ee
or conditions of the third type
\be
\label{bc3}
\reduction{ \left[\Frac{\partial}{\partial n_X}+\tau\ad(x\ad)\right]
f(X)}{\partial\Omega\bigcap\Gamma\ad}=0,
\quad x\ad\in\partial\omega\ad,\quad\ano,
\ee
on the boundary $\partial\Omega$ of the
domain $\Omega$.
Smooth functions $\tau\ad(x\ad)$
are parameters of the model and are defined for
$x\ad\in\partial\omega\ad$.
Potentials $v\ad(x\ad)$ describe pair interactions of particles
at $x\ad\!\in\!\omega\ad$ and are supposed to be smooth quickly
decreasing functions.

\section{Faddeev equations for density of simple layer }
At the beginning, we consider  the first approach \cite{MerMot},
\cite{Motovilov} to the boundary-condition model based immediately
on the Potential theory \cite{Potential}. We demonstrate it
for the case of $N\!\!=\! 3$ and conditions (\ref{bccor}) supposing also
that $v\ad\equiv 0$, $\ano$.
In this case the kernel $R(X,X',z)$, $X,X'\!\in\!\Omega,$ of resolvent
$R(z)=(H-z)^{-1}$ satisfies the identity (following from the
Green formula):
\be
\label{LSchCor}
R(X,X',z)=R_0(X,X',z)-\int_{\partial\Omega}
d\sigma_S R_0(X,S,z)
\Frac{\partial}{\partial n_S} R(X,X'z),
\ee
with $R_0(X,S,z)$, $R_0(z)=(-\Delta_X-z)^{-1}$,
the Green function\footnote{%
Remember that for $-\Delta_X$ in $L_2(\bR^\nu )$,
the Green function $\Ro(z)$
is given explicitly,
$
\Ro(X,X',z)=\Frac{i}{4} \left(\Frac{\sqrt{z}}{2\pi}\right)
\Frac{H_{(\nu-2)/2}^{(1)} (\sqrt{z}|X-X'|)}
{|X-X'|^{(\nu-2)/2}}
$,
with $H^{(1)}_{\mbox{...}}$ the Hankel function of the first type.
}
of the Laplacian $-\Delta_X$ in $L_2(\bR^{2n})$ and $n_S$ the
normal to $\partial\Omega$ directed in $\Omega$.  It follows
from the representation (\ref{LSchCor}) that  Green function
$R(z)$ is explicitly expressed in terms of its normal derivative
$\mu(Z,X'z)=\Frac{\pa}{\pa n_S} R(S,X',z)$.  The surface
integral $\Int_{\pa\Om} d\sigma_S \Ro(X,S,z)\mu(S)\equiv U(X,z)$
in (\ref{LSchCor}) is the potential of simple layer
\cite{Potential} with density $\mu$.  This potential is known to
have finite normal derivatives \cite{Potential} in all the
points $S$ where surface $\pa\Om$ is smooth,
\be
\label{DerPot}
\Lim{X\rightarrow S}\Frac{\pa U}{\pa n_S}=
\mp\Frac{1}{2}\mu(S) +\Int_{\pa\Om}\Frac{\pa}{\pa n_S}
\Ro(S,S',z)\mu(S')d\sigma_{S'}.
\ee
Here, the up (low) sign $-(+)$ corresponds to inside (outside) limit,
i.e. $X\!\in\!\Om\,$ $(X\!\in\!\bR^{2n}\setminus\overline{\Om})$.

Differentiating Equation (\ref{LSchCor}) with respect to the normal
$n_S$
and taking into account relationship (\ref{DerPot}), we obtain the
following simple-layer potential equation for $\mu(S)$:
\be
\label{IniPot}
\Frac{1}{2}\mu(S)+\Int_{\pa\Om}\Frac{\pa}{\pa n_S}
\Ro(S,S',z)\mu(S')d\sigma_{S'}=\Frac{\pa}{\pa n_S}\Ro(S,X',z).
\ee
In this equation, the variables $X'$ and $z$ are fixed parameters.
For compact integration surfaces, the potential integral equations
are known to be of Fredholm type \cite{Potential}.
Equation (\ref{IniPot}) however, is not of the Fredholm type because
the surface $\pa\Om$ is unbounded.

Let us construct the Faddeev-type \cite{Faddeev63} equations for the
density $\mu$.

First, we shall introduce some new notations. Let $\Ga\ad\re$
be the part of cylinder $\Ga\ad$ belonging to $\pa\Om$,
$\Ga\ad\re=\Ga\ad\bigcap\pa\Om$. It is clear that
$\pa\Om=\bigcup\ad\Ga\ad\re$. We shall denote by $\Ga\ad\ri$
the part of $\Ga\ad$ lying inside of $\pa\Om$,
$\Ga\ad\ri=\Ga\ad\setminus\Ga\ad\re$.
Restriction
$
\mu\ad=\reduction{\mu}{\Ga\ad\re}
$
will be called the Faddeev component of the density $\mu$.
It is convenient to consider these components as functions defined
on the total cylinders $\Ga\ad$.
Then, by $\mu\ad\re$ $(\mu\ad\ri)$
we shall denote the part of $\mu\ad(S)$ defined on $\Ga\ad\re$
($\Ga\ad\ri$). A concrete definition of  internal parts $\mu\ad\ri(S)$
of the densities $\mu\ad$ will be given in the following paragraph.

In this notation, Equation (\ref{IniPot}) may be rewritten
as the system of three coupled equations,
\be
\label{ES}
\Frac{1}{2}P\ad\re\mu\ad=P\ad\re\bV\ad\Ro - P\ad\re\bV\ad\Ro
\Sum_{\beta=1}^{3}\mu\bd\re
\ee
with $P\ad\re$ ($P\ad\ri$)  the operator of multiplication on the
characteristic function of the set $\Ga\ad\re$ $(\Ga\ad\ri)$
and $\bV\ad\Ro(z)$ the integral ope\-ra\-tor
with ker\-nel $\bV\ad\Ro(S,X',z)=$ $\Frac{\pa}{\pa n_S}\Ro(S,X',z)$,
$S\in\Ga\ad$, $X'\in\bR^{2n}$. On the other hand, we use the equations
(\ref{ES}) to define the internal parts $\mu\ad\ri$ of the densities
$\mu\ad$ replacing $P\ad\re$ with $P\ad\ri$.
Total components $\mu\ad$ satisfy the equations
$$
\Frac{1}{2}\mu\ad=\bV\ad\Ro-\bV\ad\Ro\Sum_{\beta=1}^{3}\mu\bd\re.
$$
Further, following by the Faddeev method, we transfer  diagonal terms
with $\beta=\alpha$ to the left sides and inverse operators
$\Frac{1}{2}I+\bV\ad\Ro$ appearing there. As a result we obtain the
following equations \cite{Motovilov},\cite{MerMot} for densities $\mu\ad$:
\be
\label{EFS}
\mu\ad=\bV\ad R\ad+\Frac{1}{2}\bV\rho\ad\mu\ad\ri
-\bV\ad R\ad\Sum_{\beta\ne\alpha}\mu\bd\re.
\ee
Here, $R\ad(z)$ stands for the Green function of three-body system with
only hard-core interaction in pair $\alpha$. Operator
$\bV\ad\rho\ad$ has the kernel\\
$
\bV\ad\rho\ad(S,S',z)=$ $\Int_{\Ga\ad}
\Frac{\pa}{\pa n_S}\Frac{\pa}{\pa n_{S'}}
R\ad(S,S'',z)R\ad^{N\ri}(S'',S',z) d\sigma_S
$
with
$R\ad^{N\ri}$ the Green function for the Neumann problem inside of
$\Ga\ad$.
Both kernels $\bV\ad R\ad$ and $\bV\ad\rho\ad$ are explicitly expressed
in terms of two-body subsystem $\alpha$.

The equations obtained have been studied by methods of the
Potential theory \cite{Thesis}. It was shown in particular that
after some iterations, the integral operator corresponding to
the right-hand part of (\ref{EFS}) may be present as a sum of
compact operator and another operator with norm smaller than
one. The latter is engendered by a neighborhood of the ribs of
the surface $\pa\Om$ formed by intersection of cylinders
$\Ga\ad$.  Therefore, the Fredholm alternative may be applied to
these equations and properties of the density $\mu$ may be
investigated. These properties being known, we study all the
necessary properties of the Green function $R(z)$ \cite{Thesis}.
The further procedure \cite{Thesis} of constructing the wave
operators and studying their properties (completeness,
orthogonality, asymptotics etc.) is quite analogous to that
designed for three-body problems with smooth potentials
\cite{MF}.

Equations similar to (\ref{EFS}), are obtained and studied also in the
case of the conditions (\ref{bc3}) and non-zero pair potentials $v\ad$
\cite{Thesis}.

\section{Formalism of generalized potentials}
Another approach to construct the integral as well as differential
equations for components of resolvent (and wave functions) uses
the formalism of generalized potentials \cite{MMYa},\cite{Thesis}.
Note that equation (\ref{LSchCor})
may be considered as the Lippmann-Schwinger equation with
(quasi)potential $\cV$ acting as
$
\cV f=\delta_{\pa\Om}\Frac{\pa}{\pa n}f\re.
$
Here, $\delta_{\pa\Om}\mu$ stands for generalized function
(distribution) called simple layer \cite{Vladimirov} and
$\Frac{\pa}{\pa n}f\re$, for the limit values on $\pa\Om$ (taking
from $\Om$ ) of the normal derivative $\Frac{\pa}{\pa n}f$.
Later, we shall use also notations
$\Frac{\pa}{\pa n}f\ri$ for similar limit values taken from
$\bR^{(N-1)n}\setminus\overline{\Om}$,
and $f\re$, $f\ri$ for respective limit values on $\pa\Om$ of the
function $f$ itself. The generalized function
$\delta_{\pa\Om}\mu$
acts on in accordance with the rule
$(\varphi,\delta_{\pa\Om}\mu)=\int_{\pa\Om}d\sigma_S \varphi(S)\mu(S).$
 Analogous notations will be used also in the case
when surface $\pa\Om$ is replaced with cylinders $\Ga\ad$.

Let us introduce the two-body generalized potentials $\cV\ad$,
$\cV\ad f=\delta_{\Ga\ad}\Frac{\pa}{\pa n} f\re$,
and consider instead of $H$ the new ``operator'' $\hat{H}$,
\be
\label{genS}
\hat{H}f(X)=-\Delta f(X)+\Sum_{\alpha}\cV\ad f(X),
\ee
with $X$ running all the space $\bR^{(N-1)n}$. Thereby we spread the domain
of $H$ on functions defined outside as well as inside of the surface
$\pa\Om$. According to (\ref{DerPot}) one has to suppose these functions and
their derivatives to be continuous right up to surface%
\footnote{%
Excluding the points belonging to the intersections
$\Bigcup{\alpha,\beta,\beta\ne\alpha}(\Ga\bd\bigcap\Ga\ad)$
of cylinders $\Ga\ad$, $\ano$.}
$\Ga\ad$, $\ano$.  However the breaks $f\re-f\ri\ne 0$ and
$\Frac{\pa}{\pa n}f\re-\Frac{\pa}{\pa n}f\ri\ne 0$ must be
allowed when $\Ga\ad$ is crossed. Set of these functions in
$W_2^2(\bR^{(N-1)n}\setminus\Bigcup{\alpha}\Ga\ad)$ will be
denoted by ${\cal D}$. Action of the Laplacian $-\Delta$ on
${\cal D}$ has to be understood in a sense of distributions,
\cite{Vladimirov}
\be
\label{gen}
-\Delta f=-\Delta_0 f+\Sum_\an \delta_{\Ga\ad}
\left( \Frac{\pa f\ri}{\pa n}-\Frac{\pa f\re}{\pa n} \right)+
\Sum_\an \Frac{\pa}{\pa n} \left[ \delta_{\Ga\ad}
\left(  f\ri - f\re \right) \right],
\ee
where $\Delta_0$ stands for the usual Laplacian. With (\ref{gen})
one can easily see that substitution of $\cV\ad$ into the Schr\"odinger
equation leads to the two-sided boundary conditions on $\Ga\ad$,
\eqnarray
\label{Int}
\reduction{\Frac{\pa}{\pa n}f\ri}{\Ga\ad}=0,\\
\label{Ext}
\reduction{f\ri}{\Ga\ad}=\reduction{f\re}{\Ga\ad}.
\endeqnarray
This means \cite{Thesis} that if the spectral parameter $z$ is
out of (discrete) spectrum of respective boundary value problems
for domains engendered inside of
$\bR^{(N-1)n}\setminus\overline{\Om}$ due to intersection of
cylinders $\Ga\ad$ then the spectral problem $\hat{H}\Psi=z\Psi$
for $\hat{H}$ is equivalent to that for $H$.  Therefore, the
conditions (\ref{bccor}) may be replaced outside of this
spectrum with the generalized potentials $\cV\ad$.

In the same way one can consider the model with third-type
conditions (\ref{bc3}). Generalized potentials in this case are
following~\cite{Thesis}:
$
\cV\ad f=-\delta_{\Ga\ad}(\tau\ad f\re)+
\Frac{\pa}{\pa n_{\rm e}}(\delta_{\Ga\ad} f\re)
$
(see also \cite{MMYa}).

It is easily to include in this formalism usual potentials
$v\ad$.  The only necessary thing is the replacement of $\cV\ad$
in (\ref{genS}) with $\cV\ad+v\ad$.

\section{Differential equations for components}
In the formalism of generalized potentials, the components and
differential equations for them are constructed in the same way as
in the case of usual potentials \cite{MF}.

For example, the Faddeev components of the bound-state $N$-body wave function
$\Psi$ which are defined as $U\ad=$ $-\Ro(z)[v\ad+\cV\ad]\Psi,$
satisfy the Faddeev differential equations
$$
(-\Delta_X+v\ad+\cV\ad-z)U\ad=-(v\ad+\cV\ad)\Sum_{\beta\ne\alpha}
U\bd.
$$
At $X\!\not\in\!\Bigcup{\beta}\Ga\bd,$ these equations turn into
habitual form \cite{MF}
\be
\label{dF}
(-\Delta_X+v\ad-z)U\ad=-v\ad\Sum_{\beta\ne\alpha}U\bd.
\ee
Presence of generalized potentials generates two-sided boundary
conditions for components $U\ad$ on cylinders $\Ga\ad$. In the hard-core
model (\ref{bccor}) these conditions may be written as follows:
\be
\label{bccorF}
\reduction{ \Sum_{\bn}\Frac{\pa U\bd\ri}{\pa n}  } {\Ga\ad}=0, \quad
\reduction{\Sum_{\bn} U\ad\re}{\Ga\ad}=0,
\ee
with $\alpha$ running all the numbers of pairs
(see Refs. \cite{MerMot},\cite{Motovilov},\cite{MMYa} for details).
Analogous conditions in the case of the model (\ref{bc3}) read as
\be
\label{bc3F}
\reduction{\Sum_{\bn} U\bd\ri}{\Ga\ad}=0, \quad
\reduction{\left[
\left(\Frac{\pa}{\pa n} +\tau\ad \right)
\Sum_{\bn} U\bd\re
\right] }
{\Ga\ad}=0,
\ee
In the four-body problem, we introduce after Faddeev components, those
of Ya\-ku\-bov\-sky,
$
U_{\an a}=-G\ad(v\ad+\cV\ad)\Sum_{\beta\ne\alpha,\beta\subset a}U\bd,
$
where $G\ad(z)$ is the Green function corresponding to only interaction
$v\ad+\cV\ad$. Remember that
$U\ad=\Sum_{a\supset\an} U_{\an a}$ and $\Psi=\Sum\ad U\ad$.

The components $U_{\an a}$ satisfy at $X\not\in \Bigcup{\bn}\Ga\bd$
the Yakubovsky differential equations
\be
\label{dYa}
(-\Delta_X+v\ad-z)U_{\an a}
+ v\ad\!\Sum_{ \mbox{
\scriptsize
$\begin{array}{c}
\gamma\!\ne\!\alpha,\\
\gamma\!\subset\! a
\end{array}$
}
}
\! U_{\gn a}=-v\ad\Sum_{b\ne a}
\Sum_{\mbox{\scriptsize
$\begin{array}{c}
\bn\!\ne\!\an,\\
\bn\!\subset\! a \end{array}$ }  }\! U_{\bn a}.
\ee
Boundary conditions for them may be written \cite{MMYa} in the form
\be
\label{bccorYa}
\reduction{ \Frac{\pa\Psi\ri_{\an a}}{\pa n} }{\Ga\ad}=0, \quad
\reduction{\Psi\re_{\an a}}{\Ga\ad}=0
\ee
in the hard-core model (\ref{bccor}) and
\be
\label{bc3Ya}
\reduction{\Psi\ri_{\an a}}{\Ga\ad}=0, \quad
\reduction{    \left( \Frac{\pa\Psi\re_{\an a}}{\pa n}+
\tau\ad\Psi\re_{\an a} \right)    } {\Ga\ad}  = 0
\ee
in the case of conditions (\ref{bc3}). Here, $(\an, a)$ runs all the chains
of partitions and \newline
$
\Psi_{\an a}=\Sum_{\gn\in a} U_{\gn a}+
\Sum_{\bn\ne\an}
\Sum_{\mbox{\scriptsize
$\begin{array}{c}
\bn\!\ne\!\alpha,\\
\bn\!\subset\! a          \end{array}$}    }
U_{\bn b}.
$

Components of the scattering wave functions in the boundary-condition model
satisfy the same equations (\ref{dF}) and (\ref{dYa}) and boundary
conditions (\ref{bccorF},\ref{bccorYa}) or (\ref{bc3F},\ref{bc3Ya}).
For the scattering processes, asymptotical boundary conditions
as $X\rightarrow\infty$ for the Faddeev components $(N\!\!=\! 3)$
and for the Yakubovsky components $(N\!\!=\! 4)$ are similar to those
in the case of usual potentials \cite{MF},\cite{EChAYa}. With
these conditions, the boundary value problems (\ref{dF},\ref{bccorF})
and (\ref{dYa},\ref{bccorYa}) (or (\ref{dF},\ref{bc3F})
and (\ref{dYa},\ref{bc3Ya})) become uniquely soluble (at energies
lying out of the discrete spectrum of the respective external problem
for $\pa\Om$ and some internal problems for the domains formed by
intersection of cylinders $\Ga\ad$). Results of concrete three-body
computations of $(2\rightarrow 2,3)$ scattering processes and bound-state
energies may be found in Refs.~\cite{Motovilov},\cite{EChAYa}.

\acknowledgments

Author is thankful to Prof.~\framebox[2.9cm][c]{S.P.Merkuriev} and
Dr.~S.L.Yakovlev for the fruitful collaboration and to Prof.~V.B.Belyaev
for support and active interest to this work.

Author thanks the International Science Foundation for financial
support (Project Nr.~RFB000).

\end{document}